\documentstyle[aaspptwo,epsf]{article} 

\newcommand{\um}{\mbox{~$\mu$m}}
\newcommand{\rf}{\reference}
\newcommand{\gam}{$\Gamma$}
\newcommand{\nh}{N$_H$}
\newcommand{\nhg}{N${_H,gal}$}
\newcommand{\chch}{$\chi^2$}
\newcommand{\chchr}{$\chi^2_{\nu}$}
\newcommand{\amin}{$^{\prime}$}
\newcommand{\asec}{$^{\prime \prime}$}
\newcommand{\adeg}{$^{\circ}$}
\newcommand{\secpoint}{\mbox{$''\mskip-7.6mu.\,$}}
\newcommand{\SS}[1]{\S~\ref{#1}}
\newcommand{\gapprox}{\mbox{$\stackrel {>}{_{\sim}}$}}

\newcommand{\smsub}[1]{_{\mbox{\scriptsize #1}}} 
\newcommand{\agamsub}[1]{\mbox{$\overline\Gamma\smsub{#1}$}}
\newcommand{\gamsub}[1]{\mbox{$\Gamma\smsub{#1}$}}

\begin{document}

\title{THE SOFT--X--RAY SPECTRAL SHAPE OF X--RAY--WEAK
SEYFERTS\footnote{Accepted for publication in the 10~January~1996 issue of
ApJ.}}

\author{Brian Rush and Matthew A. Malkan} \affil{Department of Physics and
Astronomy, University of California, Los Angeles, CA 90095--1562;
rush,malkan@bonnie.astro.ucla.edu}

\begin{abstract}

We present and analyze ROSAT--PSPC observations of eight Seyfert~2 galaxies,
two
Seyfert~1/QSOs, and one IR--luminous non--Seyfert. These targets were selected
from the Extended 12\um\ Galaxy Sample and, therefore, have different
multiwavelength properties from most (optically or X--ray selected) Seyferts
previously observed in the soft X--rays. The targets were also selected as
having atypical X--ray fluxes among their respective classes, e.g. relatively
X--ray strong Seyfert~2s and X--ray weak Seyfert~1/QSOs.

Comparing our observations with those from the ROSAT All--Sky Survey, we find
variability (of a factor of 1.5---2 in flux) in both of the Seyfert~1/QSOs, but
in none of the Seyfert~2s. Both variable objects have steeper photon indices in
the more luminous state, with the softest ($<$1.0~keV) flux varying the most.
The timescales indicate that the variable component arises from a region less
than a parsec in size.

Fitting the spectra to an absorbed power--law model, we find that both the
Seyfert~2s and the Seyfert~1/QSOs are best fit with a photon index of
3.1---3.2.
This is in agreement with the average photon index of a sample of Markarian
Seyfert~2s observed by Turner, Urry, \& Mushotzky (1993), indicating that most
Seyfert~2s, even those displaying a wide variety multiwavelength of
characteristics, as well as some Seyfert~1/QSOs, have a photon index much
steeper than the canonical (Seyfert~1) value of $\sim$1.7. One possible
explanation is that these objects have a flatter continuum plus a soft
($<1.0$~keV) excess in the form of high--EW iron and/or oxygen fluorescence
lines, a black--body or even a thermal plasma. Alternatively, the underlying
continuum may indeed be steep, powered by a different physical mechanism than
that which produces the flat continua in other Seyfert~1s/QSOs.

We imaged one Seyfert~2 (NGC~5005) with the ROSAT HRI, finding about 13\% of
the
soft X--rays to come from an extended source. This object also has the most
evidence from spectral fitting for an extra contribution to the soft--X--ray
flux in addition to a power--law component, indicating that different
components
to the soft X--ray spectrum of this object (and likely of other X--ray--weak
Seyferts) may come from spatially distinct regions.

\end{abstract}


\section{Introduction} \label{intro}

Although Seyfert galaxies and quasars have been well studied in the X--rays,
most previous observational scrutiny has been devoted to the brighter
Seyfert~1/QSOs which are more easily detected. There are few observations of
those Seyfert~1/QSOs which are relatively X--ray weak or of any Seyfert~2, and
not all of those have been measured well enough for detailed spectral analysis.
This paper discusses new ROSAT spectra of such objects, broadening the range of
types of AGN observed in the soft X--rays. This can provide us with an
understanding of the soft X--ray nature of (low luminosity) AGN which is more
representative of this entire class of objects, and free from the biases which
can result from analyzing only a small subset AGN types.

Previous X--ray missions, in the 2--10~keV energy range, found Seyfert galaxies
(mostly Seyfert~1s) to be best fit by power--law spectra with a photon index of
about $\Gamma\sim$1.7---1.9 (e.g. Mushotzky 1984; Turner \& Pounds 1989).
However, the ROSAT spectra of Seyferts generally have steeper photon indices,
of
about $\Gamma\sim 2.4$ for Seyfert~1s (Turner, George, \& Mushotzky 1993,
hereafter TGM) and even steeper values $\Gamma\sim 3.2$, for Seyfert~2s
(Turner,
Urry, \& Mushotzky 1993, hereafter TUM). There are several possible
explanations
for these steep observed indices. This could indicate a steeper intrinsic
continuum slope, or alternatively adding a ``soft X--ray excess" to an
underlying power--law model usually improves the fit and flattens the best--fit
continuum slope. The nature of this soft excess has been suggested to be one or
more of the following: Fe--L and/or Oxygen--K emission lines around
0.8--1.0~keV, a low--temperature blackbody, an optically--thin thermal
component, a steep second power--law, or the underlying hard continuum leaking
through a partial absorber. It is not evident that a combination of a
power--law
and a soft excess is necessary in all objects. Perhaps a large amount of
absorption ($N_H\sim10^{23}$) could harden an even softer underlying power--law
to give the observed spectrum, or a strong blackbody or optically--thin thermal
component could account for all of the observed soft--X--ray flux, without an
underlying power--law even being necessary.

These large object--to--object differences in the observed range of
$L_x/L_{opt}$ in Seyfert~1s and QSOs of a factor of 300 (e.g., values of
$\alpha_{ox}$ ranging from --1.0/--1.1 to --1.9---Picconotti et al. 1982;
Tananbaum et al. 1986) reflect substantial fundamental differences in the
structure of their central engines. A large difference in X--ray properties is
also seen in the spectra of Seyfert~2s. For example, NGC~1068, the prototype of
a Seyfert~2 which may be a hidden Seyfert~1, is also the brightest and best
observed Seyfert~2 in the X--rays. It appears to have a very steep soft X--ray
spectrum (Monier \& Halpern 1987), but is more like Seyfert~1s at high energies
(Koyama et al. 1989), and does not resemble the average spectrum of other
Seyfert~2s observed with the IPC, or the spectrum of the Seyfert~2 Mkn~348
observed with Ginga (Warwick et al. 1989).

These differences, lead to the question of whether the usual
Seyfert~1---Seyfert~2 dichotomy, usually made based on optical spectra, is a
physically accurate way to classify these objects in the X--rays. Observations
of a wide range of Seyfert galaxies are necessary to determine whether
Seyfert~1s and Seyfert~2s represent two primarily distinct classes of objects,
or if they are better described as having a continuous {\it range\/} of
properties, and whether the observed differences are intrinsic to the nucleus,
or represent varying circumnuclear properties, such as the amount and
distribution of absorbing material. Our data suggest that a subset of
Seyfert~1s
(of which we discuss only two objects in this work, but which may include many
other objects) are more intrinsically similar (with respect to the source of
the
soft X--ray emission) to most Seyfert~2s than to other Seyfert~1s. This is most
likely explainable if different mechanisms produce the X--rays in the
X--ray--quiet objects. If the standard X--ray emission mechanisms (e.g.,
inverse--Compton scattering of lower energy photons by relativistic electrons,
direct synchrotron emission from relativistic electrons produced near the
central engine or jet, and/or thermal emission from the hot inner parts of an
accretion flow) are in fact virtually ``turned off" in these objects, it is
quite possible that weaker, more exotic mechanisms (e.g., optically thin
thermal
emission from the hot intercloud medium) may contribute significantly to the
X--rays we actually detect.

\section{Target Selection and Observations} \label{targets}

\subsection{Selection of Objects from the 12 Micron Sample}
\label{targets_selection}

The objects for which we have obtained pointed PSPC spectra were carefully
selected for several reasons. First, they are from (with the exception of
PG~1351+640) the most complete and unbiased source of bright AGNs compiled to
date---the Extended 12 Micron Galaxy Sample (Rush, Malkan, \& Spinoglio 1993).
This sample is complete relative to a {\it bolometric\/} flux level, and
includes those Seyferts which are the brightest at longer wavelengths,
including
a truly representative number of both X--ray--quiet and X--ray--loud objects.
We
selected the IR--brightest Seyfert~2s from this sample which had not previously
been observed in any pointed X--ray mission. We also selected two typical
examples of relatively X--ray--weak Seyfert~1/QSOs. Mkn~1239 has one of the
lowest detected X--ray fluxes of all 55 Seyfert~1s in the 12\um\ Sample (20
counts and 0.05~cts/sec in the ROSAT All--Sky Survey---Rush et al. 1996), and
PG~1351+640 has the steepest $\alpha_{ox}$ (-1.91) of the 66 PG~~QSOs observed
by Einstein (Tananbaum et al. 1986).

Second, the 12\um--selected Seyferts are qualitatively different from those
observed previously. Halpern \& Moran (1993) pointed out that the Seyfert~2s
usually observed, with polarized broad lines, are restricted to those with
relatively strong UV excesses (found by the Markarian surveys; e.g. those
reported in TUM) which are also relatively radio--strong. Compared to these
Markarian Seyfert~2s (many of which were observed but not detected by
Ginga---Awaki 1993), the targets we observed have redder optical/infrared
colors, weaker and smaller radio sources, larger starlight fractions, and
steeper Balmer decrements---more representative of Seyfert~2s as a general
class. Similarly, Mkn~1239 and PG~1351+640 differ from those broad--line AGN
usually observed, in that they are specifically chosen to have relatively weak
X--ray fluxes. The one IR--luminous non-Seyfert we observed was chosen by
cross--referencing the non--Seyferts in the 12\um\ Sample with a large sample
of
IRAS galaxies detected in the ROSAT All--Sky Survey (hereafter RASS; Boller et
al. 1992; Boller et al. 1995b) for those non--Seyferts with the highest IR
luminosity {\it and\/} X--ray flux.

\subsection{Pointed ROSAT PSPC Observations during AO2--AO4}
\label{targets_obs}

The observations were carried out AO2---AO4 (from 1991~December to
1993~October)
with the ROSAT X--ray telescope, with the Position Sensitive Proportional
Counter (PSPC) in the focal plane. The PSPC provides spatial and spectral
resolution over the full field of view of 2\adeg\ which vary slightly with
photon energy E. The energy resolution is $\Delta$E/E = 0.41/$\sqrt{E_{keV}}$.
The on--axis angular resolution is limited by the PSPC to about 25\asec, and
the
on--axis effective collecting area, including the PSPC efficiency, is about
220~cm$^2$ at 1~keV (Brinkmann 1992). See Table~1 for a summary of the
observations and count rates for each object, where the objects are listed in
decreasing order of total counts obtained.

We have also obtained ROSAT All--Sky Survey data for almost all of the Seyferts
in the 12\um\ and CfA samples. This will be discussed in another paper to be
completed shortly after this one (Rush et al. 1996). Those data, on over 100
Seyferts spanning a wide range of characteristics, will complement this work by
enabling us to address {\it statistically\/} the scientific issues discussed
below for individual objects.

\section{Data Analysis} \label{analysis}

For each step of the data analysis discussed below, only those counts in pulse
invariant (PI) channels 12---200 inclusive are included. The lower limit is set
by the fact that the lower level discriminator lies just below this limit, so
any data taken from lower channels cannot be considered as valid events.
Furthermore, analysis of the PSPC PSF has shown that the positions of very soft
events cannot be accurately determined because of a ghost imaging effect (J.
Turner, p.comm).  The exact level at which this effect is significant is
different for each observation (Hasinger \& Snowden 1990), so we conservatively
chose to exclude PI channels below 12. The upper PI channel included is 200,
since the mirror effective area falls off rapidly at higher energies. We have
also defined low, medium, and high energies to refer to PI channels 12---50,
51---100, and 101---200, respectively, and ``all" energies refers to PI
channels
12---200.

The spectral analysis was done by first extracting spectra from the events file
using the QPSPEC command in the PROS package in IRAF. We made sure that the
output of PROS were properly compatible with XSPEC, in particular with regards
to the manner in which these two packages deal with binning and calculating
statistical errors.\footnote{This simple but very important procedure is
explained in detail at http://heasarc.gsfc.nasa.gov/docs/rosat/to\_xspec.html.}
We then fit simple models using the XSPEC software, with the events in PI
channels 12---200 binned so as to include at least 20 counts in each bin,
allowing \chch\ techniques to be applied.\footnote{We only required 10 counts
per bin both NGC~3982 and CGCG~022--021, and 5 counts per bin in NGC~1144, in
order to have at least 7 bins for the fits; this makes the results extremely
rough, but otherwise we would have only 3--4 bins, with which no fits could be
done.} We used the most recent response matrix available, released from MPE in
1993~January. We first fit the data to the standard absorbed power--law model,
both with all parameters (\gam, \nh, and normalization) free and with \nh\
fixed
at the Galactic value (see Table~2). We use the photon index, \gam, defined
such
that $N_\nu \propto \nu^{-\Gamma}$ ($N$ = number of photons), which is output
by
the fitting routines in XSPEC. This relates to the spectral slope, $\alpha$,
defined by $F_\nu \propto \nu^{\alpha}$, as $\Gamma = 1 - \alpha$. We also
performed several other fits, either adding a thermal component to the
power--law or fitting only a thermal component. These are discussed in
\SS{results_fits}

The quoted uncertainties are at the 90\% confidence level, assuming one free
parameter of interest (Lampton, Margon, \& Bowyer 1976), when available (i.e.,
when the chi--square minimization to determine these uncertainties properly
converged; these are denoted as separate upper and lower uncertainties).
Otherwise, the 1$\sigma$ uncertainty on each parameter is given (denoted as a
single $\pm$ value.)

Hardness ratios provided a simple approximation to the spectral shape, even for
those objects which didn't have enough counts to accurately fit a spectral
model
to (see Table~3). The hardness ratio is defined as HR=(A--B)/(A+B), where
A~=~ctrt~(0.12--1.00~keV) and B~=~ctrt~(1.01--2.00~keV). Also given is the
ratio
A/(A+B), which we refer to as F$_{\mbox{soft}}$.

The spatial analysis was done using the SAOimage display in IRAF/PROS. Each of
the sources were observed at the center of the PSPC field, with the exception
of
NGC~1144, which was about 20\amin\ south of the field center. This object was
partially occulted by the telescope support structure and we thus corrected the
exposure time accordingly. The accumulated PSPC counts for each object were
calculated using the IMCNTS task in IRAF/PROS and are listed in Table~1. All
counts in a circular region surrounding the source are given, after subtracting
the background, as calculated in a source--free annular region just outside the
circle.

Finally, using the TIMSORT and LITCURV tasks in PROS, we extracted light curves
for each object. This was done individually for low, medium, and high energies
and for all energies. All of the objects were observed over periods of no more
than 8 days, except for NGC~3982 and PG~1351+640, which were observed in
several
segments, spanning 5 and 11 months, respectively, allowing us to test for
variations on a half--year to year time scale.

\section{Results} \label{results}

\subsection{Variability} \label{results_var}

\subsubsection{Seyfert~2s}

Any variation in the spectra of our Seyfert~2s would have be considered an
important result, as there are only a couple reports to date of X--ray
variability in Seyfert~2 galaxies (e.g., in NGC~1365---TUM and, possibly, in
Mkn~78---Canizares et al. 1986), and none of these are conclusive (e.g., the
variation in NGC~1365 may be due to the serendipitous sources). However, no
significant short--term variation was found for any Seyfert~2 in our sample.
The
one object which was observed over a 5~month period, NGC~3982, showed no
significant variation over this time scale either (see, for example, the count
rates in Table~1).

We also compared the count rates of our pointed observations to those obtained
during the ROSAT All--Sky Survey for the same objects (Rush et al. 1996), as
shown in Figure~1. Point sizes in Figure~1 are proportional to the square of
the
total counts\footnote{Several figures have point sizes proportional to counts
instead of count--rate or flux. This is because the former is also an indicator
of SNR and thus also of the statistically accuracy of spectral fits and other
quantitative results. Also, this makes little difference since the exposure
times vary only by a factor of two among our objects while the total counts
vary
by a factor of $\sim$20.} in our pointed observation and errorbars are
$1\sigma$
statistical uncertainties in the count rates. The RASS was taken during
1990~July---1991~February, thus this comparison provides timelines of 1---3
years for the various objects. As can be seen, the 5 Seyfert~2s with the most
counts in our observations show no sign of variability since the RASS. That the
count rates for two of the fainter Seyfert~2s and for the one IR--luminous
non--Seyfert are different is probably {\it not\/} an indication of
variability,
since we have extremely low counts for those objects (in both our observations
and the RASS), and it is unlikely that only the objects with the fewest
observed
count rates would be the only ones to vary.

\subsubsection{Seyfert~1/QSOs}

However, there {\it is\/} evidence for variation in both of our Seyfert~1/QSOs.
{}From Table~1 and Figure~1, we can see that Mkn~1239 increased its count rate
by
about a factor of two between the RASS and our observation (over 21---28
months,
depending on when this object was observed during the RASS). The spectral slope
steepened slightly during this period, from $\Gamma=2.69$ to $\Gamma=2.94$ (for
a power--law fit, with \nh\ constrained to \nhg, which is the only spectral
parameter we have from the RASS).

We don't have RASS data for PG~1351+640, but we can see that it varied during
our observations, which spanned the 11 months from 1992~November to
1993~October, increasing its total counts and flux by factors of 1.5 and 1.4,
respectively (a $\sim10\sigma$ result). The spectral shape varied, becoming
steeper as this object became more luminous, as with Mkn~1239. The
0.12---1.00~keV count rate increased by $\sim$59\%, whereas the 1.00---2.00
count rate only increased by $\sim$14\%, as indicated by the counts and
hardness
ratios of Table~3. The best--fit photon index steepened slightly, from 2.54 to
2.73 (see Table~2).

That the spectra of both of these objects steepened during the more luminous
state indicates that most of the variability was at the lowest energies (i.e.,
below 1~keV). The timescale of the variability puts an upper limit on the size
of the emitting region for this soft component, of much less than a light--year
for PG~1351+640, and less than two light--years for Mkn~1239, restricting the
source to the area not much larger than the broad--line region.

\subsection{Spectral Fitting} \label{results_fits}

\subsubsection{Power--Law Models} \label{results_fits_pl}

We fit each of our spectra to a simple absorbed power--law model, both with
\nh\
held constant at the Galactic value, and allowing it to vary. As an example, we
show in Figure~2 the data and folded model for our highest SNR object,
PG~1351+640. Below we discuss how the spectra for the other objects differ. We
also show, in Figure~3, the \chch\ contour plot which results from minimizing
\chch\ as a function of \nh\ and \gam\ for this object. The contours represent
the 68\%, 90\%, and 99\% confidence limits (1$\sigma$, 1.6$\sigma$, and
2.6$\sigma$, respectively) and the plus marks the best--fit value. The contour
plots for our strongest 6 objects (in terms of total counts---PG~1351+640;
NGC~5005; Mkn~1239; NGC~424; NGC~4388; and NGC~5135) look roughly the same as
this one, and those for the other objects look increasingly ``bent", with less
well--defined maxima as the total number of photons decreases.

As indicated in Table~2, when \nh\ is allowed to vary, the best--fit value is
always higher than the Galactic value, by a factor of 2---3 (again, for the 6
well--determined spectra), the one exception being PG~1351+640 which shows no
increase. The fact that \chchr\ (reduced \chch) decreases by $\sim$35-50\% when
allowing \nh\ to vary indicates that these values are more accurate than the
Galactic ones. This indicates that there is indeed some internal absorption of
one form or another in these objects, and that the underlying slope is steeper
than that which is obtained when requiring \nh=\nhg. We illustrate this in
Figure~4, where we plot the photon indices obtained with \nh\ free versus with
\nh\ fixed. Most of our Seyfert~2s, as well as those from TUM, have the former
steeper by $\sim1$.

The average values of \gam\ which we obtain with \nh\ free are
$\overline\Gamma=3.13$ for our 4 Seyfert~2s with sufficient counts, and
$\overline\Gamma=3.20$ for our two Seyfert~1/QSOs. These values are similar to
the six Seyfert~2s observed by TUM, which have $\overline\Gamma=3.16$, but
differ from the six Seyfert~1/QSOs observed by TGM which have
$\overline\Gamma=2.41$.

In Figure~5, we plot the photon index versus count rates for the pointed
observations of this work, TUM, and TGM. We see that most of the objects have
significantly steeper values of \gam\ than the old canonical value of 1.7
(dotted line). All of our well--observed Seyfert~2s (filled triangles), and
most
of TUM's Seyfert~2s (open triangles), {\it and\/} both of our Seyfert~1/QSOs
have values of $\Gamma\sim3$. The one exception is Mkn~372 which has a value of
$\Gamma=2.2$. However this object is now known to be a Seyfert~1, and, as
expected lies close to the average value of the Seyfert~1/QSOs from TGM at
$\overline\Gamma\sim2.4$.

What these data show us is that, not only do most Seyfert~2s have a best--fit
photon index around $\Gamma\sim3$, but also that Seyfert~1s are divided between
objects which have similar spectral slopes as Seyfert~2s and those which have
flatter spectra with $\Gamma\sim2.2$. Physical explanations for this are
discussed further in \SS{disc} and~\SS{summary}

\subsubsection{Internal Absorption} \label{results_fits_abs}

For each of our targets, we looked at the best--fit hydrogen column density as
compared to the Galactic value, and compared this to the photon indices and
hardness ratios, to try to determine the significance of internal absorption
and
how this affects the observed count rates and spectral shape. Figure~5 seems to
indicate that a few of the faintest objects also have the hardest spectra. This
is tentative, however, since these objects are the ones with the fewest photons
and the data are not very trustworthy. However, we do note that, if real, this
is consistent with these faint objects being the most heavily absorbed (i.e.,
with low signal--to--noise, a heavily absorbed, intrinsically steep spectrum
would appear similar to a relatively unabsorbed flat spectrum). We investigate
this trend further by plotting the spectra of our 8 brightest objects in
Figure~6 (in order of brightness, from the upper left, down to the lower
right),
fit to a power--law with \nh\ free. The general trend is for the fainter
objects
to have harder spectra (as also indicated by the hardness ratios in Table~3),
with the 4 highest hardness ratios belonging to 4 of the 5 lowest--count
objects
(the exception being NGC~3982 which actually has one of the lowest hardness
ratios).

To determine whether these harder--spectrum objects may be more heavily
obscured
by dust, we have compared their ROSAT hardness ratios to their IRAS colors (see
Figure~7). Six of our objects are very dusty in the far--IR, having values of
$\log F_{\nu,60} / F_{\nu,25}$ $\sim0.8-1.0$, which is among the reddest
(which probably means most dust--enshrouded) third of even Seyfert~2s (Rush et
al. 1993). This includes the four lowest--count objects in our sample.
Conversely, both PG~1351 and Mkn~1239 have values of $\log F_{\nu,60} /
F_{\nu,25}$ $\sim0.15$, which is among the hottest $\sim$20\% of even
Seyfert~1s. However, there is no strong relation of the IRAS color to the
hardness ratio other, other than that of the three hardest objects are also
among the reddest.

Taken together, these results indicate that there is a trend for the fainter
objects to have harder ROSAT spectra, indicating that absorption is partially
responsible for steepening the spectra. However there is less evidence that the
amount of absorption is correlated with redness/dustiness in the galaxy, as
determined from IRAS colors.

\subsubsection{Additional Models} \label{results_fits_other}

We also fitted some of our spectra to other models. These include a power--law
plus an emission line or thermal component (Raymond--Smith thermal plasma or
blackbody), or a thermal component alone. As discussed in \SS{individual} for
individual objects, there are several cases where the fits improve, indicating
that more than a simple power--law may be necessary to explain the soft
X--rays.

First, we added an additional component to the underlying power--law. The fits
to neither of our Seyfert~1/QSOs were improved by adding another component.
This
is as expected, as the power--law fits to both objects were quite good (\chchr\
of 0.79 and 0.67 for PG~1351+640 and MKN~1239, respectively). The fit did
improve, however when we added an emission line to some of our Seyfert~2s. See,
for example, Figure~8 which shows the model for a power--law plus gaussian
emission line fit to NGC~5005. The best--fit energy for this line is at
0.8~keV,
around the energy expected for Fe--L and/or Oxygen--K emission lines. Adding
this component also has the effect of flattening the underlying power--law
slope
from 3.0 to 2.4. Similar results are obtained for the fits to NGC~5135 and
NGC~4388, which are slightly improved by adding emission lines at 0.5, and
0.6~keV, respectively.

We also tried fitting each object to a thermal model only. Again, both
Seyfert~1/QSOs were not fit at all well in this way. However, several
Seyfert~2s
(NGC~5005, NGC~5135, NGC~5929, and NGC~1144), were fit better (i.e., lower
\chch\ for the same number of free parameters) by a $\sim$0.2~keV black--body
than by an absorbed power--law (see, for example Figure~9 for the black--body
fit to NGC~5135). This is significant in that it prevents us from saying
conclusively that the soft--X--rays from these objects are associated with the
AGN at all, and that they may simply be due to stellar processes. It is not
likely that ROSAT data alone will be able to finally distinguish between
stellar
and non--stellar explanations for the X--ray emission from Seyfert~2s, as the
most definitive tests to discriminate between such models are best done in the
hard X--rays (e.g., Iwasawa 1995).

\subsection{Spatial Extent} \label{results_extent}

\subsubsection{HRI Image of NGC~5005} \label{results_extent_hri}

If multiple components are responsible for the soft--X--rays in these objects,
it is quite possible that they are from spatially distinct regions, as is
already known to be the case for some brighter Seyfert galaxies. For example,
the brightest and best observed Seyfert~2 in the X--rays is NGC~1068, the
prototype of a Seyfert~2 which may be a hidden Seyfert~1. HRI Imaging (Wilson
1994; Halpern 1992; Wilson et al. 1992) of this object reveals at least three
components to the soft--X--ray emission: (a) a compact nuclear source,
coincident with the optical nucleus, (b) asymmetric emission extending
10--15\asec\ N---NE, closely correlated with the radio jet and narrow--line
[OIII] emission, and (c) large--scale (60\asec) emission with similar
morphology
to the starburst disk. These three components comprise 55, 23, and 22\% of the
X--ray flux, respectively.

To investigate whether similar structures may be responsible for part of the
soft X--rays from our (much fainter) objects, we obtained a 27~ksec HRI
exposure
of our brightest Seyfert~2 galaxy, NGC~5005, shown in the contour plot in
Figure~10 (the contour values range from 0.05 to 0.60 photons/pixel and the
spatial resolution is 0\secpoint5/pixel). The central source spans
$\sim$20\asec~x~20\asec, and is significantly extended (FWHM$\sim$10\asec) as
compared to the HRI on--axis PSF (FWHM$\sim$5\secpoint5). The position of the
peak of this central component agrees within error to the optical position, and
is roughly 3\secpoint7 south of the radio--interferometer position given by
Vila
et al. (1990).

In addition to this central component, there is an extended wing from about
10\asec\ to 25\asec\
to the south--west of the central source (from 0.6$h^{-1}$~kpc
to 1.4$h^{-1}$~kpc). This feature contains about 13\% as many
background--subtracted counts as does the central source (31 compared to 247).
The orientation of this feature is roughly parallel to the major optical axis
of
the galaxy ($\sim45^\circ$ E of N), although the latter represent structure on
the 1--arcminute scale. At smaller sizes, arcsecond--scale radio maps made with
the VLA at 6~and 20~cm are presented in Vila et al. (1990). They find the
central source to dominate the nuclear region of the galaxy (being marginally
resolved---FWHM$\sim$0\secpoint7), and weak extended structure over
$\sim$2~arcsec in no particular direction.


Although this is our brightest Seyfert~2 galaxy, the spatial resolution and
counts are only sufficient to tell that there definitely is some asymmetric
soft--X--ray emission. Higher spatial--resolution and higher SNR data of
X--ray--weak Seyferts with future X--ray missions will be necessary to
determine
the general significance of the contribution of extended components to the
soft--X--ray spectrum of such objects.

\subsubsection{PSPC Images} \label{results_extent_pspc}

None of targets show extended emission in the PSPC image. (However, not being
primarily an imaging instrument, the resolution of the PSPC would only show
structure on much larger scales than the HRI, and cannot be used to rule out
sub--arcminute--scale structure, as exemplified by the fact that our HRI image
of NGC~5005 clearly shows structure not apparent in the PSPC images of the same
object.) Several of the images contain field objects $\sim$10--20\amin\ from
the
target, clearly distinguished by the resolution of the PSPC. The only exception
is NGC~1144, which is not spatially separated from NGC~1143. Since the latter
is
a non--active galaxy the X--rays are likely to be mostly from NGC~1144, however
we note the PSPC spectrum is a combination of these two sources.\footnote{This
object has the least counts of all, primarily due to obscuration by the
telescope support structure, so no strong conclusions can be drawn about its
spectrum.} It is interesting to note that TUM found serendipitous (optically)
unidentified X--ray sources about 1\amin\ from each of the six Seyfert~2s
observed in their program. In some cases (e.g., NGC~1365) these sources are
likely bright X--ray sources in the host galaxy, and in others (e.g., Mkn~78)
they are likely low--luminosity AGNs. We looked for such sources in the field
of
our 12\um\ Seyfert~2s, and found none. The number of Seyfert~2s (14) observed
between these two samples makes it highly unlikely that this difference could
be
explained simply by chance. One possible explanation is that the objects in TUM
are galaxies previously known to be relatively bright in the X--rays from
Einstein IPC observations, and these serendipitous sources could have
contributed to the Einstein flux.

\section{Discussion} \label{disc}

\subsection{The Standard Soft X--Ray Slope for X--Ray Weak Seyferts}
\label{disc_newslope}

Considering both our data and that of TUM, it appears that a steep spectral
slope, around \gam=3, should be considered the standard slope for X--ray--weak
Seyferts. This includes virtually all Seyfert~2s, as indicated by the results
that have been derived for Seyfert~2s displaying a wide range in
multiwavelength
characteristics. As discussed in \SS{targets_selection}, our objects were
chosen
from the 12\um\ sample and thus have redder optical/infrared colors than the
objects observed by TUM, which are Markarian objects selected as having a
strong
UV--excess.

Even the prototypical Seyfert~2 galaxy, NGC~1068, resembles these objects.
Monier \& Halpern (1987) observed this object with Einstein, finding a
0.1---3.8~keV photon index of $\Gamma\sim3.0$, and \nh\ consistent with the
Galactic value. Our data from the RASS give a 0.1---2.0~keV value of
$\Gamma=2.78$ for this object (Rush et al. 1996), which is slightly harder, but
consistent when considering that our RASS data was fitted with \nh\ constrained
to \nhg.

This category of X--ray--steep AGN not only includes most Seyfert~2s, but some
X--ray--weak Seyfert~1 /QSOs, such as PG~1351+640 and Mkn~1239. That the soft
X--ray source in these objects may be the same as in most Seyfert~2s is
consistent with their selection as being X--ray {\it weak\/} for
Seyferts~1/QSOs. In contrast, other Seyfert~1/QSOs, e.g. those observed by TGM,
were known to be relatively strong in the soft X--rays, and thus one would
expect those objects to have X--ray spectra more similar to conventional
Seyfert~1s. Thus, it seems that the standard Seyfert~2---Seyfert~1 dichotomy in
not the simplest way to categorize these AGN in the soft X--rays. Rather, we
could refer to (relatively) steep, X--ray--weak objects and flat,
X--ray--strong
objects, whose soft X--rays are probably dominated by different components.

We also find steep average spectral slopes in our RASS data (to be analyzed
thoroughly in Rush et al. 1996), of \agamsub{Sy1}=2.24$\pm0.49$ and
\agamsub{Sy2}=2.86$\pm0.48$ for 39 Seyfert~1s and 5 Seyfert~2s, respectively
(uncertainties quoted are 1$\sigma$ individual scatter). These fits were done
with \nh\ constrained to \nhg, and thus the best--fit slopes are likely a
little
steeper, depending mainly on the amount of internal obscuration. This could
place the average slope of the Seyfert~2s over 3 and that of the Seyfert~1s
around 2.4---2.5. This and the fact that there is a wide range of slopes for
the
Seyfert~1s, with over 1/3 being steeper than \gam=2.5 assuming no internal
absorption, makes these results consistent with those for our pointed
observations---namely that all Seyfert~2s and some Seyfert~1s have slopes much
closer to 3 than to 2. Similar results have been found in other works, for
example Boller, Brandt, \& Fink (1995a), who surveyed 46 narrow--line
Seyfert~1s
with ROSAT and found them all to have extremely steep spectra (some with \gam\
as high as 5).

\subsection{Physical Interpretation} \label{disc_interpretation}

There are several competing explanations for the steep slopes observed in many
X--ray--weak Seyferts, as compared to the flatter slopes observed in
conventional (X--ray--strong) Seyferts. The physical models which may be able
to
explain all or part of the observed differences between steep--slope and
flat--slope Seyferts include:

(1) A separate, hard power--law present in steep objects which is very weak,
such as a scattered component. Although we see no evidence of such a component
in our fits, we cannot rule out this possibility, as observations in a larger
wavelength baseline of X--ray--weak Seyferts may detect such a component if it
is extremely faint.

(2) Much of the soft spectrum of steep objects being produced by the same
physical mechanism, located in the same place, as the soft excess observed in
many flat objects. In this model, steep objects have relatively more soft
excess
and less of the hard power--law.

The evidence for this type of spectrum would be that fits to a power--law--only
model would give a very steep slope, but that adding the soft excess would
flatten the underlying slope while improving the fit. As discussed in
\SS{results_fits_other} and \SS{individual}, we have evidence for this in
several of our objects, and even a pure black--body with no underlying
power--law cannot be ruled out in some cases. This is even more evident in TUM,
as most of their objects are fitted significantly better when either an
emission
line or Raymond--Smith plasma are added to the power--law. If we do assume that
a very soft excess exists in these objects, a physical model for this excess
still remains to be determined. For example, it could be thermal emission from
the galaxy, hot gas near the nucleus, iron and/or oxygen emission line(s), or
the UV bump shifted into the ultra--soft X--rays as suggested in Boller et al.
(1995a). But, again, we stress that such evidence is not universal, as several
of our objects show no definite preference for anything other than a
power--law.

(3) That the soft spectrum we see in X--ray--weak Seyferts represents a
component present in most or all Seyferts, but which is much weaker in X--ray
strong objects and is thus suppressed by the hard spectrum in those objects. If
so, is this universal component non--nuclear, i.e. similar to the soft X--rays
observed in normal or starburst galaxies (from, e.g., X--ray binaries and
SNRs)?

(4) That the soft spectra arise from the same physical process (and from the
same location) as the flat power--laws in some Seyfert~1s, but with a higher
value for \gam, caused by variance of one or more intrinsic physical
parameters?
For example, of several explanations Boller et al. (1995a) suggest for their
steep spectra, one of the more promising ones is that the central engine in
these objects is at a lower mass than other Seyfert~1s, and would thus have an
accretion disk emitting at a higher temperature, shifting the UV bump into the
low--energy end of the ROSAT band, steepening the X--rays. This idea is also
one
possible explanation for the steep spectra we found in PG~1351+640 and
MKN~1239,
as well as other X--ray--weak Seyfert~1/QSOs. To test this idea thoroughly, one
would need to observe the {\it spread\/} in \gam\ for many X--ray--weak and
X--ray--strong Seyferts and see if there is a continuous range of observed
values, as opposed to a more--or-less bimodal distribution. If such a range is
observed, then determining any X--ray or multiwavelength parameter which is
correlated with \gam\ would provide information about the fundamental cause of
its variance.

Finally, an important caveat in this distinction between X--ray--weak and
strong
Seyferts is that our X--ray--weak Seyfert~1/QSOs are not exactly like our
Seyfert~2s in the soft X--rays, which is seen in several ways: (1) even though
the former have the same steep slope when fitted to a power--law, they are more
often fitted only by this steep power--law, as opposed to a power--law plus an
additional component (and PG~1351+640 cannot be fitted at all by any model
other
than a pure power--law); (2) they are also more luminous in the soft X--rays
than all but the very strongest Seyfert~2s; and (3) they show less indication
of
internal absorption (above the Galactic value): of all our objects, PG~1351+640
is the only one to not have even the slightest evidence for internal absorption
in a power--law fit, and several of our Seyfert~2s show much stronger evidence
for internal absorption than does MKN~1239. This last difference is of
particular importance because it can affect the measured parameters in each of
the models listed above. These differences imply that, although the observed
soft X--ray emission from these Seyfert~1/QSOs is similar to that from
Seyfert~2s, the underlying physical processes are probably at least partially
different. Perhaps, for example, the X--ray--weak Seyfert~1/QSOs are best
explained by one or more of the models listed above, but the Seyfert~2s by
another. Thus, whereas is seems as though these relatively X--ray weak
Seyfert~1/QSOs should definitely not be strictly grouped with the more luminous
(flat--slope) Seyfert~1/QSOs with regards to the soft X--ray properties, they
still appear somewhat distinct from even the relatively X--ray strong
Seyfert~2s
and perhaps represent an intermediate or mixed class.

\section{Notes on Spectral Fits to Individual Objects} \label{individual}

\subsection{PG~1351+640 and Mkn~1239}

These two Seyfert~1/QSOs were relatively well observed, with 990 and 595 counts
obtained, respectively. Both were well fitted with a simple power--law. For our
strongest object, PG~1351+640, no improvement is obtained by allowing \nh\ to
vary, giving no indication of internal absorption. For Mkn~1239, an increase of
about a factor of 1.5 in \nh\ over the Galactic value reduces \chchr\ from 0.95
to 0.67, perhaps indicating some internal absorption.

We tried to fit each object to the other models listed in Table~2. For
PG~1351+640, the parameters returned each time indicated that a single
power--law was preferred (i.e., the normalization for other component was at or
near zero). Mkn~1239, on the other hand, fit well to a power--law model with
the
addition of a gaussian emission line around 0.7~keV. This fit was not, however
better than those with a Raymond--Smith plasma or black--body replacing the
emission line. Thus, if there is a second component to the soft X--rays
spectrum, we cannot distinguish among several possibilities for its shape.

For PG~1351+640, we also separately fit the spectra which were taken during
1992~November and 1993~October to a power--law model. A slight increase in the
best--fit \gam\ is found in the more luminous state.

\subsection{NGC~424, NGC~4388, NGC~5005, and NGC~5135}

These four Seyfert~2s each yielded at least $\sim$400 counts (see table~1),
sufficient for accurate spectral fitting. For these objects, an average photon
index of $\overline\Gamma=3.13$ (3.0, 3.2, 3.2, and 3.2, respectively) was
obtained when \nh\ was allowed to vary, and of $\overline\Gamma=2.00$ (1.7,
2.1,
1.9, and 2.3) when \nh\ was constricted to the Galactic value.

In all cases, we tried adding another component to the fit. In the case of
NGC~5135 the fit was improved at a significance level of $>90\%$. This object
has the hardest spectrum of these four Seyfert~2s. Considering that it is also
fitted by the largest \nh, the hard spectrum and the good fit to a second
component above 0.5~keV both probably indicate significant absorption of the
softest X--rays below 0.5~keV. Adding emission lines also improved the fits to
NGC~5005 ($>99\%$ significance level) and NGC~4388 ($>90\%$). Only in the case
of NGC~5005 was the emission line at the energy expected for Fe--L and/or
Oxygen--K, thus identification of these components with a specific emission
process is not possible. We also fit NGC~5005 and NGC~5135 to a black--body
model and obtained better fits than to a power--law model, further indicating
that we don't know the source of the soft X--rays---whether they are from the
nonstellar active nucleus or from stellar processes such as X--ray binaries or
supernova. In the latter case, we have some evidence that a small contribution
of the soft--X--rays may come from an extended component, as discussed in
\SS{results_extent_hri} for NGC~5005.

\subsection{IRAS~F01475--0740 and NGC~5929}

For these two objects, only 276 and 200 counts were obtained, allowing only 12
and 9 points (bins) for the spectral fitting, respectively. Interestingly,
relative to the 0.5---2.0~keV range, F01475--0740 has almost no counts below
0.5~keV, and NGC~5929 has very few. In fact, F01475--0740 has the hardest
spectrum of any object we observed, indicted both by the hardness ratios in
Table~3 and by the very flat value of \gam. NGC~5929 also has a harder spectrum
than any of the objects discussed above, but not nearly as hard as
F01475--0740.
This may indicate that these objects are very heavily absorbed, which would
explain both the low overall flux and the hard spectra.

When adding another component to the power--law for F01475--0740, \gam\ always
tended towards zero (as flat as we would allow), with only a small contribution
from the other component---indicating nothing more than the very hard spectrum
of the simple power--law. For NGC~5929, a slight improvement in the fit was
obtained by adding a second component, similar to some of the brighter four
Seyfert~2s discussed above, but with much less statistical significance.

\subsection{NGC~3982 and NGC~1144}

These two objects yielded so few counts that can only give a very rough
estimate
of the best--fit photon index, which is 2.12 and 1.90 for NGC~3982 and
NGC~1144,
respectively with \nh\ fixed. Only NGC~3982 had enough photons to allow a fit
with \nh\ variable, which yielded $\Gamma=3.4$. Although this slope is similar
to the values for our bright Seyfert~2s, the spectra do not look similar.
NGC~3982 has the softest and NGC~1144 the second hardest count rates of any of
our Seyfert~2s. There were not enough counts to fit to composite models, but we
did try to fit these spectra to a simple black--body, to estimate whether or
not
a power--law is even the most descriptive of the soft X--rays. For NGC~3982
there was only marginal improvement in the fit, but for NGC~1144 \chchr\ did
drop by almost a factor of two for the black--body fit as compared to a
power--law.

\subsection{CGCG~022--021}

In addition to the 10 Seyfert galaxies discussed above, we also observed one
IR--luminous non--Seyfert which had been detected by the ROSAT All--Sky Survey.
We would expect the ROSAT spectra of this type of object to be similar to those
from Seyfert~2s (both of which emit strongly in the thermal infrared, but
relatively weakly in the X--rays), if the X--ray emission in the latter are
produced by the normal processes of stellar evolution, as in classic starburst
nuclei like NGC~7714 (Weedman et al. 1981).

Unfortunately, the observation of CGCG~022--021 yielded only 81$\pm$30 counts,
and a count--rate of 0.010 $\pm$0.003 cts/s, which is not sufficient for a
detailed spectral analysis. There may be some indication of variability, since
the RASS count--rate was 0.064 $\pm$0.018 cts/s, indicating a $\gapprox2\sigma$
change. However, this is very tentative as the (background--subtracted) counts
obtained in the pointed and RASS observations are only 81 and 26, respectively.

We do see, though, that this non--seyfert has a hard spectrum quite similar to
that several of the weaker Seyfert~2s (F01475--0740, NGC~5929, and NGC~1144).
This indicates that heavy internal absorption is probably present. To describe
the spectrum further, we attempted to fit simple models to the X--ray flux,
although with high uncertainties. A simple power--law and a black--body model
provided similarly accurate fits (\chchr\ of 1.2 and 1.3, respectively),
however
the error bars are high.

\section{Summary and Conclusions} \label{summary}

We have analyzed pointed ROSAT PSPC spectra of 11 objects selected as having
atypical soft X--ray fluxes. These include 8 Seyfert~2s and one IR--luminous
non--Seyfert selected from the Extended 12\um Galaxy Sample, which all have
relatively strong detections in the ROSAT All--Sky Survey, as compared to other
objects in their class. We also observed on X--ray weak Seyfert~1/QSO from this
sample and a similar object selected from the PG Bright Quasar Survey.

We found both Seyfert~1/QSOs, Mkn~1239 and PG~1351+640, to vary in flux by a
factors of 2 and 1.5, over periods of less than 2 and 1 year, respectively.
Both
objects had steeper spectra in their more luminous state, indicating that the
variability was mainly due to the softest X--rays, which are confined to a size
of less than a parsec.

All of our Seyfert~2s which had sufficient counts for accurate spectral
fitting,
as well as both Seyfert~1/QSOs, have soft X--ray photon indices of $\sim3$,
similar to the Seyfert~2s observed by TUM. The wide--spread occurrence of such
steep slopes suggests that this value of $\Gamma\sim3$ is the norm for a wide
variety of AGN, namely Seyfert~2s {\it and\/} many Seyfert~1/QSOs. Therefore,
discussing relatively steep ($\Gamma\sim3$), X--ray--weak objects versus flat
($\Gamma\sim2$), X--ray--strong objects may be a more fundamental way to
separate
Seyferts with respect to the soft X--rays than the usual type~1--type~2
dichotomy (derived primarily from optical spectra).

There are several possible explanations for these steep slopes. One is the
presence of a very soft ($<1$~keV) excess in addition to a flatter underlying
continuum. We see strong evidence in the spectral fits to some of our objects
for such a component, but a physical model for this excess still needs to be
determined---it could be strong iron and/or oxygen line emission, a
black--body,
or even a thermal plasma. However, several of our objects show no definite
preference for anything other than a steep power--law. Alternatively, both flat
and steep components could be present in some Seyferts, with one or the other
dominating depending on internal physical conditions. Or the steep and flat
spectra observed in different objects may have the same basic origin, but with
variance of one or more parameters affecting the measured slope. Distinguishing
between these and other models for the X--ray emission from Seyferts can best
be
done by testing multiple--component models over the entire 0.1---10~keV range,
where the distinguishing spectral signatures of competing models can be most
clearly identified. Thus, obtaining high---SNR spectra of X--ray weak Seyferts,
with several thousand of counts both in the soft and hard X--rays, should prove
a profitable pursuit of current and future X--ray missions.

Finally, we obtained a ROSAT HRI image of one Seyfert~2 (NGC~5005) and found
about 13\% of the flux to come from an extended component. This implies that
multiple components of the soft--X--ray spectra of Seyferts may arise in
spatially distinct regions, as has been previously observed primarily in
brighter objects. Further, deeper images of X--ray--weak Seyferts will be
necessary to determine the physical processes giving rise to these components,
as well as how common such phenomena are in Seyfert galaxies.

\acknowledgements

We thank Jane Turner for much help in understanding the PROS and XSPEC
software,
the ROSAT data, and the specifications of the PSPC, and for providing us with
the results of TUM and TGM before publication. This work was supported by NASA
grants NAG~5--1358 and NAG~5--1719.

\clearpage

\addtocounter{page}{+3} 

\clearpage

\centerline{\bf FIGURE LEGENDS}

\noindent {\bf Figure 1} --- Our pointed PSPC count rates versus count rates
from the ROSAT All--Sky Survey. Squares are Seyfert~1/QSOs, triangles are
Seyfert~2s, and the star is our IR--luminous non--Seyfert. $\mbox{Point
sizes}\propto\sqrt{\mbox{total counts}}$. Error bars are $1\sigma$ statistical
uncertainties. The solid line represents CTRT$_{\mbox{\scriptsize Pointed}}$ =
CTRT$_{\mbox{\scriptsize RASS}}$.

\noindent {\bf Figure 2} --- PSPC Spectrum of PG~1351+640, fit to an absorbed
power--law with \nh\ free.

\noindent {\bf Figure 3} --- \chch\ contour plot of \nh\ vs. \gam\ for the fit
shown in Figure~2. Contours represent confidence limits of 68, 90, and 99\% and
the plus marks the best--fit value.

\noindent {\bf Figure 4} --- Photon Index for power--law fits: with \nh\ free
versus \nh\ constrained to \nhg. The solid lines represent
\gamsub{free}~=~\gamsub{gal} and \gamsub{free}~=~\gamsub{gal}~+~1. Symbols are
the same as in Figure~1, with open triangles representing Seyfert~2s from TUM.

\noindent {\bf Figure 5} --- Photon Index for power--law fits with \nh\ free,
versus log count rate. Symbols are the same as in Figure~1, with the addition
of
open squares and open triangles for the Seyfert~1/QSOs in TGM and the
Seyfert~2s
in TUM, respectively. $\mbox{Point sizes}\propto\sqrt{\mbox{total counts}}$.
The
dotted line shows the canonical value of $\Gamma=1.7$. For the Seyfert~1/QSOs
from TGM, there was little spread in \gam\ (5 of 6 objects between 2.11---2.50
and the other---Mkn~335---at 3.10), and thus only the average value is shown
here.

\noindent {\bf Figure 6} --- PSPC spectra of all of our 8 brightest objects,
each fit to an absorbed power--law with \nh\ free.  The objects are placed in
order of total counts obtained, starting with PG~1351+640 in the upper left,
going down each column, to NGC~1144 in the lower right. \\ (Figure~6 is Placed
LAST among the figures.)

\noindent {\bf Figure 7} --- IRAS 25---60\um\ color versus hardness ratio.
Symbols sizes are proportional to total counts.

\noindent {\bf Figure 8} --- Model of the fit of a power--law plus emission
line
to our PSPC spectrum of NGC~5005, where the individual components are shown.
The
dot--dash line is a gaussian emission line at 0.8~keV, the long dashed line is
the absorbed power--law, and the solid line is the total model.

\noindent {\bf Figure 9} --- PSPC Spectrum of NGC~5135, fit to a black body
model.

\noindent {\bf Figure 10} --- Contour plot made from our 27~ksec HRI Image of
NGC~5005. Contours range from 0.05 to 0.60 photons/pixel. The spatial
resolution
is 0\secpoint5 per pixel.

\end{document}